\documentclass[aps,twocolumn,groupedaddress,amsmath,amssymb]{revtex4}

\usepackage{graphicx}
\usepackage{dcolumn}
\usepackage{xcolor}%
\usepackage{booktabs}%
\usepackage{bm}

\begin{document}

\title{Superconductivity with large upper critical field in noncentrosymmetric Cr-bearing high-entropy alloys}

\date{\today}
\author{Guorui Xiao$^{1,2,3}$}
\author{Wuzhang Yang$^{1,2,4}$}
\author{Qinqing Zhu$^{1,2,4}$}
\author{Shijie Song$^{3}$}
\author{Guang-Han Cao$^{3}$}
\author{Zhi Ren$^{1,2}$}
\email{renzhi@westlake.edu.cn}

\affiliation{$^{1}$Department of Physics, School of Science, Westlake University, 18 Shilongshan Road, Hangzhou, 310024, Zhejiang Province, PR China}
\affiliation{$^{2}$Institute of Natural Sciences, Westlake Institute for Advanced Study, 18 Shilongshan Road, Hangzhou, 310024, Zhejiang Province, PR China}
\affiliation{$^{3}$School of Physics, Zhejiang University, Hangzhou 310058, PR China}
\affiliation{$^{4}$Department of Physics, Fudan University, Shanghai, 200433, PR China}

\begin{abstract}
A series of new Cr$_{5+x}$Mo$_{35-x}$W$_{12}$Re$_{35}$Ru$_{13}$C$_{20}$ high-entropy alloys (HEAs) have been synthesized and characterized by x-ray diffraction, scanning electron microscopy, electrical resistivity, magnetic susceptibility and specific heat measurements. It is found that the HEAs adopt a noncentrosymmetric cubic $\beta$-Mn type structure and exhibit bulk superconductivity for 0 $\leq$ $x$ $\leq$ 9.
With increasing $x$, the cubic lattice parameter decreases from 6.7940(3) {\AA} to 6.7516(3) {\AA}. Meanwhile, the superconducting transition temperature $T_{\rm c}$ is suppressed from 5.49 K to 3.35 K due to the magnetic pair breaking caused by Cr moments.
For all these noncentrosymmetric HEAs, the zero-temperature upper critical field $B_{\rm c2}$(0) is comparable to Pauli paramagnetic limit $B_{\rm P}$(0) = 1.86$T_{\rm c}$. In particular, the $B_{\rm c2}$(0)/$B_{\rm P}$(0) ratio reaches a maximum of $\sim$1.03 at $x$ = 6, which is among the highest for $\beta$-Mn type superconductors.
\end{abstract}

\maketitle
\maketitle

\begin{figure*}
	\includegraphics*[width=17cm]{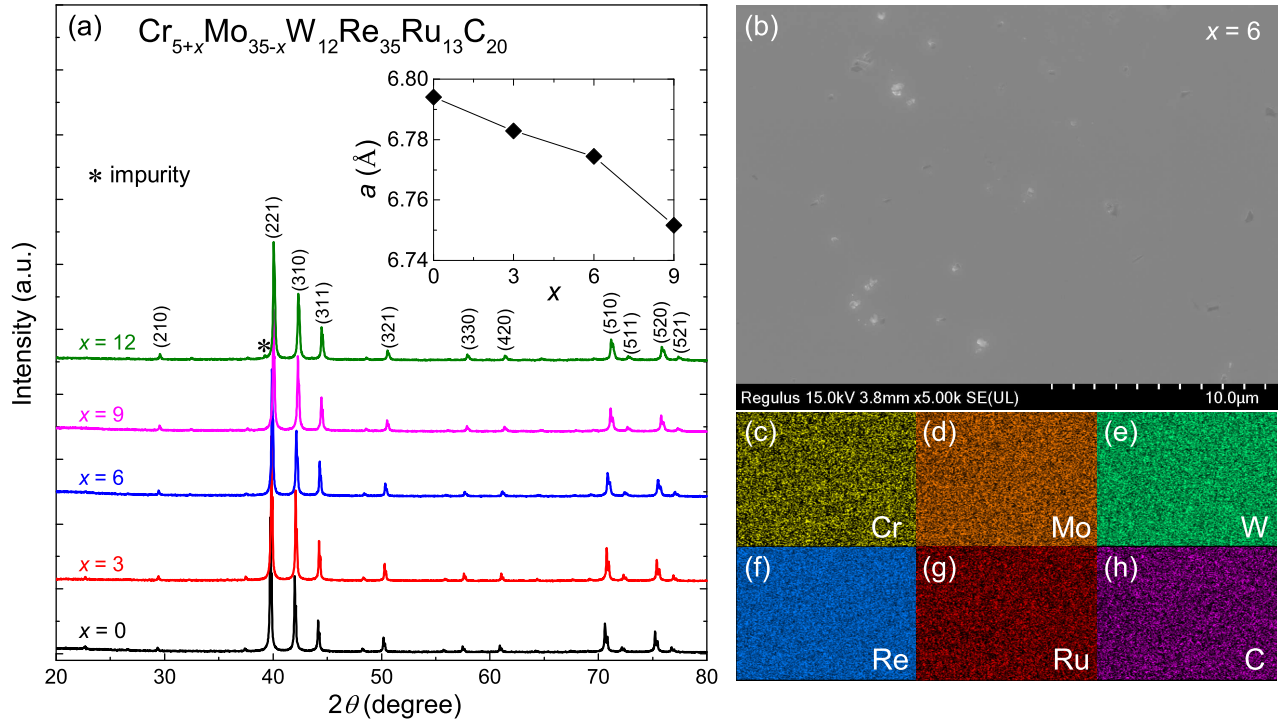}
	\caption{
		(a) XRD patterns for the Cr$_{5+x}$Mo$_{35-x}$W$_{12}$Re$_{35}$Ru$_{13}$C$_{20}$ HEAs with increasing $x$ from 0 to 12.
         For $x$ = 12, the major diffraction peaks are indexed and the impurity peak is marked by the asterisk.
         The inset shows the refined cubic lattice parameter plotted as a function of $x$.
		(b) SEM image for the HEA with $x$ = 6.
        (c-h) EDX elemental maps for this HEA.
	}
	\label{fig1}
\end{figure*}
Recently, the design of alloys based on the high-entropy concept has attracted widespread attention \cite{yeh2004nanostructured,ye2016high,miracle2017critical,zhang2018science,george2019high,li2021mechanical,wang2021high}. In principle, high-entropy alloys (HEAs) should be single phase solid solutions made up of at least five metallic elements whose atomic concentrations vary between 5\% to 35\%.
These materials are believed to be stabilized by the large configurational entropy and characterized by strong atomic disorder and local lattice distortion. As such, HEAs are commonly refereed as metallic glasses on ordered lattices and often exhibit superior mechanical \cite{lee2020lattice,an2021novel}, chemical \cite{tang2014alloying,zhang2014microstructures,shi2017corrosion,hua2021mechanical,vaidya2019phase,whitfield2021assessment} and physical \cite{yuan2017rare,law2021increased,von2016effect,guo2017robust,sun2019high} properties compared with traditional alloys based on one or two elements.
Apart from this, the chemical complexity of HEAs allows for a wide tunability in both structure and property,
which offers a fertile playground to study their relationship in multicomponent alloy systems.

So far, a number of HEAs have been found to display type-II superconductivity and most of them possess centrosymmetric structural types, such as
body-centered-cubic ($bcc$) type \cite{kovzelj2014discovery,marik2018superconductivity}, CsCl type \cite{stolze2018sc}, hexagonal-closed packed ($hcp$) type \cite{marik2019superconductivity,lee2019superconductivity,liu2020superconductivity}, A15 type \cite{wu2020polymorphism,yamashita2021synthesis}, $\sigma$ type \cite{liu2020formation,liu2021superconductivity} and face-centered-cubic ($fcc$) type \cite{zhu2022structural}.
The $T_{\rm c}$ of these HEAs shows a peculiar dependence on the valence electron concentration and is robust again disorder and magnetic impurity.
By contrast, the absence of structural inversion symmetry can allow for a mixed spin-singlet and spin-triplet pairing state.
However, noncetrosymmetric HEA superconductors have been much less explored and exist only in cubic $\alpha$-Mn \cite{stolze2018high,liu2021structural} and $\beta$-Mn \cite{xiao2022centrosymmetric} type structures. The former examples include (ZrNb)$_{1-x} $(MoReRu)$_{x}$, (HfTaWIr)$_{1-x} $Re$_{x}$ \cite{stolze2018high}, (HfTaWPt)$_{1-x} $Re$_{x}$ \cite{stolze2018high}, Nb$_{25}$Mo$_{5+x}$Re$_{35}$Ru$_{25-x}$Rh$_{10}$ \cite{liu2021structural}, while the latter example is limited to Ta$_{10}$Mo$_{5}$W$_{30}$Re$_{35}$Ru$_{20}$C$_{x}$ (16 $\leq$ $x$ $\leq$ 20) \cite{xiao2022centrosymmetric}, where the interstitial carbon plays a crucial role in stabilizing the structure. Moreover, to our knowledge, no prior study has been conducted to look for noncentrosymmetric HEA superconductors containing magnetic elements.

Here we present the synthesis and characterization of new Cr$_{5+x}$Mo$_{35-x}$W$_{12}$Re$_{35}$Ru$_{13}$C$_{20}$ HEAs.
For the first time, a single $\beta$-Mn type phase is obtained for $x$ in the range of 0 to 9.
These $\beta$-Mn type HEAs become bulk superconductors with $T_{\rm c}$ decreasing monotonically from 5.49 K to 3.35 K as the increase of $x$, which is attributed to pair breaking due to Cr local moments.
Despite the suppression of $T_{\rm c}$, the $B_{\rm c2}$(0) for all these HEAs is close to or even slightly exceeds the corresponding $B_{\rm P}$(0).
A comparison is made between the $B_{\rm c2}$(0)/$B_{\rm P}$(0) ratios of Cr$_{5+x}$Mo$_{35-x}$W$_{12}$Re$_{35}$Ru$_{13}$C$_{20}$ HEAs and ternary $\beta$-Mn type superconductors, and its implication is briefly discussed.

Polycrystalline Cr$_{5+x}$Mo$_{35-x}$W$_{12}$Re$_{35}$Ru$_{13}$C$_{20}$ HEAs with 0 $\leq$ $x$ $\leq$ 12 were prepared by the arc-melting method.
Stoichiometric amounts of high purity Cr (99.95\%), Mo (99.9\%), W (99.9\%), Re (99.99\%), Ru (99.9\%) and C (99.5\%) powders were mixed thoroughly and pressed into pellets in an argon filled glove-box. The pellets were then melted four times in an arc furnace under high-purity argon atmosphere, followed by rapid cooling on a water-chilled copper plate. The crystal structure of resulting samples was characterized by powder X-ray diffraction (XRD) using a Bruker D8 Advance X-ray diffractometer with Cu K$\alpha$ radiation. The lattice constant was determined by the Lebail fitting using the JANA2006 programme \cite{petvrivcek2014crystallographic}. The morphology and chemical composition were investigated in a Hitachi Regulus 8230 field emission scanning electron microscope (SEM) equipped with an energy dispersive X-ray (EDX) spectrometer. Electrical resistivity and specific heat measurements were done in a Quantum Design Physical Property Measurement System (PPMS-9 Dynacool). DC magnetization measurements were carried out in a Quantum Design Magnetic Property Measurement System (MPMS3).

The XRD patterns for the series of Cr$_{5+x}$Mo$_{35-x}$W$_{12}$Re$_{35}$Ru$_{13}$C$_{20}$ HEAs are displayed in Fig. 1(a).
For $x$ varying from 0 to 12, the major diffraction peaks look similar and can be well indexed on the noncentrosymmetric cubic $\beta$-Mn type structure with the $P$4$_{1}$32 space group.
This is confirmed by structural refinement, whose result for $x$ = 6 is shown in Supplementary Fig. S1.
Note that all the transition metal atoms are assumed to be distributed randomly in the lattice.
Hence the configurational entropies are calculated to be 1.40$R$, 1.45$R$, 1.49$R$ and 1.51$R$ ($R$ = 8.314 J mol$^{-1}$ K$^{-1}$ is the molar gas constant) for $x$ = 0, 3, 6, and 9, respectively, which are listed in Table I.
Nonetheless, a small impurity peak is observed near the (221) peak for $x$ = 12, and hence the remaining of the paper is focused on the HEAs with 0 $\leq$ $x$ $\leq$ 9.
The inset of Fig. 1 (a) shows the refined lattice constants plotted as a function of $x$.
One can see that the $a$-axis decreases from 6.7940(3) {\AA} to 6.7516(3) {\AA} as the increase of $x$.
This is as expected since the atomic radius of Cr (1.267 {\AA}) is smaller than that of Mo (1.386 {\AA}) \cite{pauling1947atomic}.
\begin{figure*}
	\includegraphics*[width=16cm]{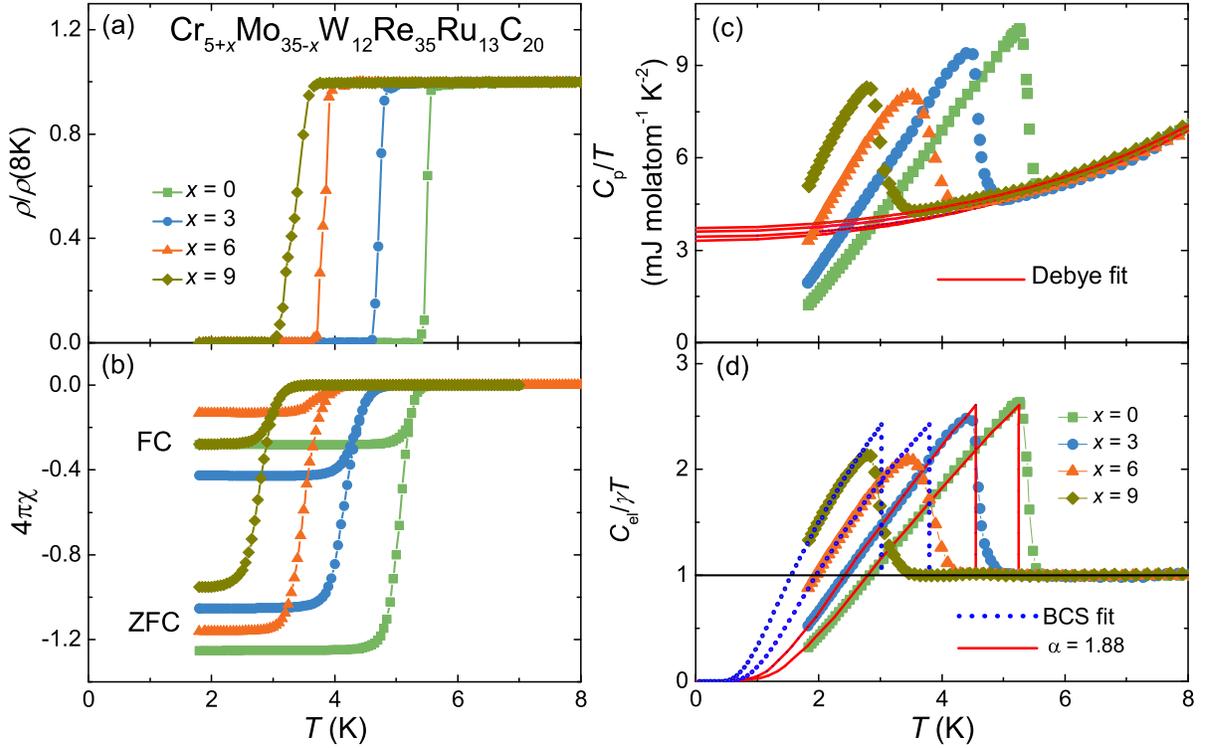}
	\caption{
		(a-c) Low temperature normalized resistivity, magnetic susceptibility and specific heat, respectively, for the series of Cr$_{5+x}$Mo$_{35-x}$W$_{12}$Re$_{35}$Ru$_{13}$C$_{20}$ HEAs.
         In panel (c), the solid lines are Debye fits to the normal-state data.
          (d) Temperature dependence of normalized electronic specific heat for the HEAs. The dashed and solid lines are fits to the data by the BCS theory and $\alpha$-model, respectively.
	}
	\label{fig2}
\end{figure*}

\begin{figure*}
	\includegraphics*[width=17.7cm]{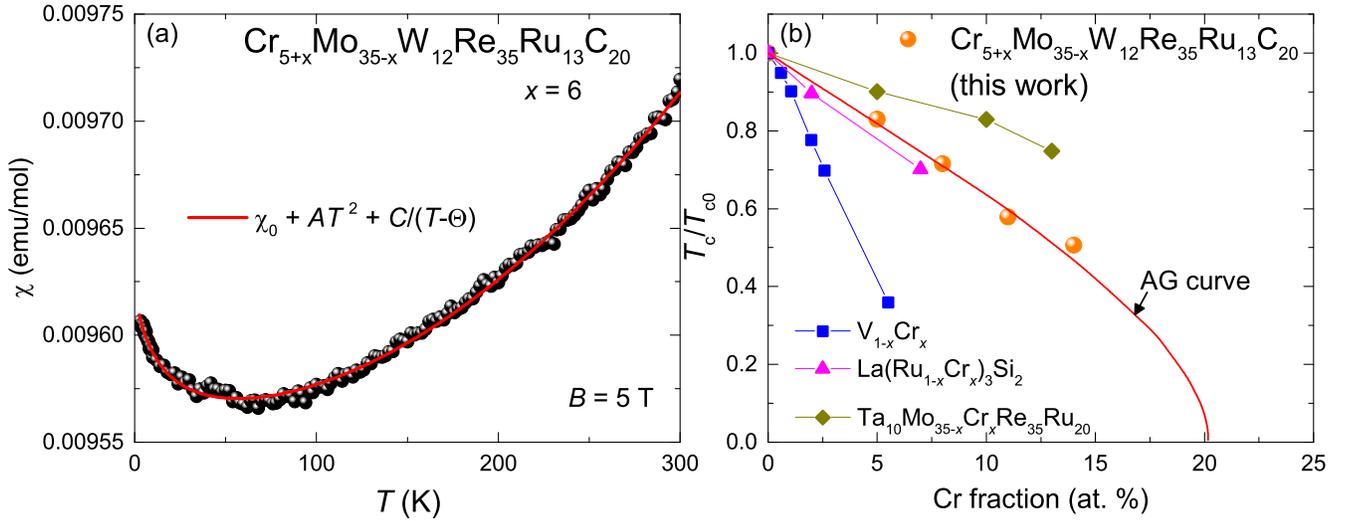}
	\caption{
	(a) Temperature dependence of normal-state magnetic susceptibility for the Cr$_{5+x}$Mo$_{35-x}$W$_{12}$Re$_{35}$Ru$_{13}$C$_{20}$ HEA with $x$ = 6. The solid line is a fit to the data (see text for details). (b) Dependence of normalized $T_{\rm c}$ on the Cr atomic fraction for the  Cr$_{5+x}$Mo$_{35-x}$W$_{12}$Re$_{35}$Ru$_{13}$C$_{20}$ HEAs. The solid line denotes the curve predicted by the AG pair-breaking theory. The data for typical elemental and intermetallic superconductors are also included for comparison.
	}
	\label{fig3}
\end{figure*}

\begin{table*}
	\caption{Parameters of the Cr$_{5+x}$Mo$_{35-x}$W$_{12}$Re$_{35}$Ru$_{13}$C$_{20}$ HEAs.}
	\renewcommand\arraystretch{1.3}
	\begin{tabular}{p{4cm}<{\centering}p{2.5cm}<{\centering}p{2.5cm}<{\centering}p{2.5cm}<{\centering}p{2.5cm}<{\centering}}
		\\
		\hline
		Parameter          & $x$ = 0 & $x$ = 3 & $x$ = 6 & $x$ = 9 \\
		\hline
		{$a$ ({\AA})}& {6.7940(3)}& {6.7828(2)}&{6.7745(1)}&{6.7516(3)}\\
	    {Configurational entropy}  & 1.40$R$& 1.45$R$& 1.49$R$& 1.51$R$\\
		{Cr (at.\%)}  &  4.7(1)& 8.1(1)& 11.1(1)& 13.3(1)\\
	  	{Mo (at.\%)}  &  33.5(1)& 31.3(1)& 28.5(1)& 25.4(1)\\
		{W (at.\%)}   &  12.1(1)& 11.4(1)& 11.8(1)& 11.0(1)\\
		{Re (at.\%)}  &  35.8(1)& 34.9(1)& 34.7(1)& 35.7(1)\\
		{Ru (at.\%)}  & 13.9(1)& 14.3(1)& 13.9(1)& 14.6(1)\\
		$T_{\mathrm{c}}$ (K) & 5.49& 4.73  & 3.83 & 3.35 \\
		$\gamma$ (mJ molatom$^{-1}$ K$^{-2}$) & 3.31& 3.44 & 3.60& 3.72 \\
		$\delta$ (mJ molatom$^{-1}$ K$^{-4}$) & 0.048 & 0.041 & 0.039 & 0.039  \\
		$\Theta_{\mathrm{D}}$ (K)   & 343& 362   & 368& 368 \\
		$\lambda_{\mathrm{ep}}$      & 0.63 & 0.60 & 0.56 & 0.55 \\
        $B_{\rm c2}$(0) (T)     & 9.7& 8.4  & 7.3& 5.3 \\
        $B_{\rm c2}$(0)/$B_{\rm P}$(0)     & 0.95& 0.96  & 1.03& 0.85 \\
		$\xi_{\mathrm{GL}}(0)$ (nm)   & 5.8& 6.3 & 6.7& 7.9 \\
		\hline
	\end{tabular}
	\label{Table1}
\end{table*}

Fig. 1(b) shows a typical SEM image for the Cr$_{5+x}$Mo$_{35-x}$W$_{12}$Re$_{35}$Ru$_{13}$C$_{20}$ HEA with $x$ = 6.
Except for a few carbon particles with size below 1 $\mu$m, the HEA appears to be dense and homogeneous.
This is corroborated by the EDX elemental mapping results displayed in Fig. 1(c)-(h), which reveal a homogeneous distribution of all the constituent elements.
In addition, as listed in Table I, the ratios of Cr:Mo:W:Re:Ru are determined to be 4.7:33.5:12.1:35.8:13.9, 8.1:31.3:11.4:34.9:14.3, 11.1:28.5:11.8:34.7:13.9, and 13.3:25.4:11.0:35.7:14.6 for $x$ = 0, 3, 6, and 9, respectively.
It is obvious that the measured metal stoichiometries are in agreement with the nominal ones within the experimental error.
As for carbon, its content cannot be determined accurately by the EDX method due to its light mass.

The low temperature normalized resistivity ($\rho$), magnetic susceptibility ($\chi$) and specific heat ($C_{\rm p}$) data for the series of Cr$_{5+x}$Mo$_{35-x}$W$_{12}$Re$_{35}$Ru$_{13}$C$_{20}$ HEAs are displayed in Fig. 2(a)-(c).
For all $x$ values, a rapid $\rho$ drop, a large diamagnetic $\chi$ and a clear $C_{\rm p}$ jump are detected, indicating the occurrence of bulk superconductivity.
From the midpoint of $\rho$ drop, $T_{\rm c}$ is determined to be 5.49 K, 4.73 K, 3.83 K and 3.35 K for $x$ = 0, 3, 6, and 9, respectively.
Below $T_{\rm c}$, there is a obvious divergence between the zero-field cooling (ZFC) and field cooling (FC) $\chi$ curves, which is characteristic of a type-II superconducting behavior.
At 1.8 K, the zero-field cooling $\chi$ data correspond to shielding fractions, $-$4$\pi$$\chi$, in the range of 95-125\% without demagnetization correction.
In the normal state, the $C_{\rm p}$ data can be well fitted by the Debye model,
\begin{equation}
C_{\rm p}/T = \gamma + \beta T^{2} + \delta T^{4},
\end{equation}
where $\gamma$ and $\beta$($\delta$) are the electronic and phonon specific heat coefficients, respectively. The obtained $\gamma$ and $\beta$ values are listed in Table 1.
Once $\beta$ is known, the Debye temperatures $\Theta_{\rm D}$ are calculated to be 343 K, 362 K, 368 K and 368 K for $x$ = 0, 3, 6, and 9, respectively, according to the equation,
\begin{equation}
\Theta_{\rm D} = (12\pi^{4} R/5\beta)^{1/3}.
\end{equation}
 Then the electron-phonon coupling strength $\lambda_{\rm ep}$ is found to decrease from 0.63 to 0.55 with increasing $x$ from 0 to 9 based on the inverted McMillan formula \cite{PhysRev.167.331},
\begin{equation}
\lambda_{\rm ep}=\frac{1.04+\mu^{*} \ln \left(\Theta_{\rm D} / 1.45 T_{\mathrm{c}}\right)}{\left(1-0.62 \mu^{*}\right) \ln \left(\Theta_{\rm D} / 1.45 T_{\mathrm{c}}\right)-1.04},
\end{equation}
where $\mu^{\ast}$ = 0.13 is the Coulomb repulsion pseudopotential.

Fig. 2(d) shows the normalized electronic specific heat $C_{\rm el}$/$\gamma$$T$ obtained by subtraction of the phonon contribution.
For $x$ = 6 and 9, the $C_{\rm el}$/$\gamma$$T$ jump follows nicely the weak-coupling BCS theory \cite{PhysRev.108.1175}.
In comparison, the $C_{\rm el}$/$\gamma$$T$ jumps become larger at $x$ $\leq$ 3.
Hence, to analyze the data, the so called $\alpha$-model \cite{johnston2013elaboration} is employed.
This model still assumes a fully isotropic superconducting gap but allows for the variation of coupling constant $\alpha$ $\equiv$ $\Delta$(0)/$k_{\rm B}$$T_{\rm c}$, where $\Delta$(0) is the gap size at 0 K.
It turns out that the $C_{\rm el}$/$\gamma$$T$ jumps for both $x$ = 0 and 3 can be well reproduced with $\alpha$ = 1.88.
This $\alpha$ value is indeed larger than $\alpha_{\rm BCS}$ = 1.764 of BCS theory \cite{PhysRev.108.1175}, consistent with the trend in $\lambda_{\rm ep}$.
Overall, these results suggest that the Cr$_{5+x}$Mo$_{35-x}$W$_{12}$Re$_{35}$Ru$_{13}$C$_{20}$ HEAs are weakly coupled, fully gapped superconductors.

Given that Cr is a magnetic element, it is natural to investigate the normal-state magnetic behavior of the Cr$_{5+x}$Mo$_{35-x}$W$_{12}$Re$_{35}$Ru$_{13}$C$_{20}$ HEAs.
Fig. 3(a) shows the $\chi$($T$) curve measured under 5 T for the HEA with $x$ = 6.
With decreasing temperature from 300 K, $\chi$($T$) decreases gradually but exhibits a upturn below 70 K.
Notably, the whole curve is well fitted by the equation \cite{PhysRevB.68.132507}
\begin{equation}
\chi = \chi_{0} + AT^{2} + \frac{C}{T-\Theta},
\end{equation}
where $\chi_{0}$ is the temperature independent term, $AT^{2}$ is the temperature dependent Pauli paramagnetism (due to the thermal depopulation of electronic states at the Fermi level), $C$ is the Curie constant, and $\Theta$ is the Weiss temperature. The obtained $\chi_{0}$ = 9.55$\times$10$^{-3}$ emu mol$^{-1}$, $A$ = 1.78$\times$10$^{-9}$ emu mol$^{-1}$ K$^{-2}$, $C$ = 1.06$\times$10$^{-3}$ emu K mol$^{-1}$, and $\Theta$ = $-$15.5 K. Assuming that the $C$ term is solely due to the Cr atoms, it gives an effective moment of 0.28 $\mu_{\rm B}$ per Cr atom.
The $\chi$($T$) behavior is very similar for other $x$ values, and applying the same analysis gives Cr effective moments varying between 0.25 and 0.54 $\mu_{\rm B}$ (see Supplementary Fig. S2).
These values are more than half or comparable to that in Cr metal \cite{arrott1967neutron}.
\begin{figure*}
	\includegraphics*[width=16.8cm]{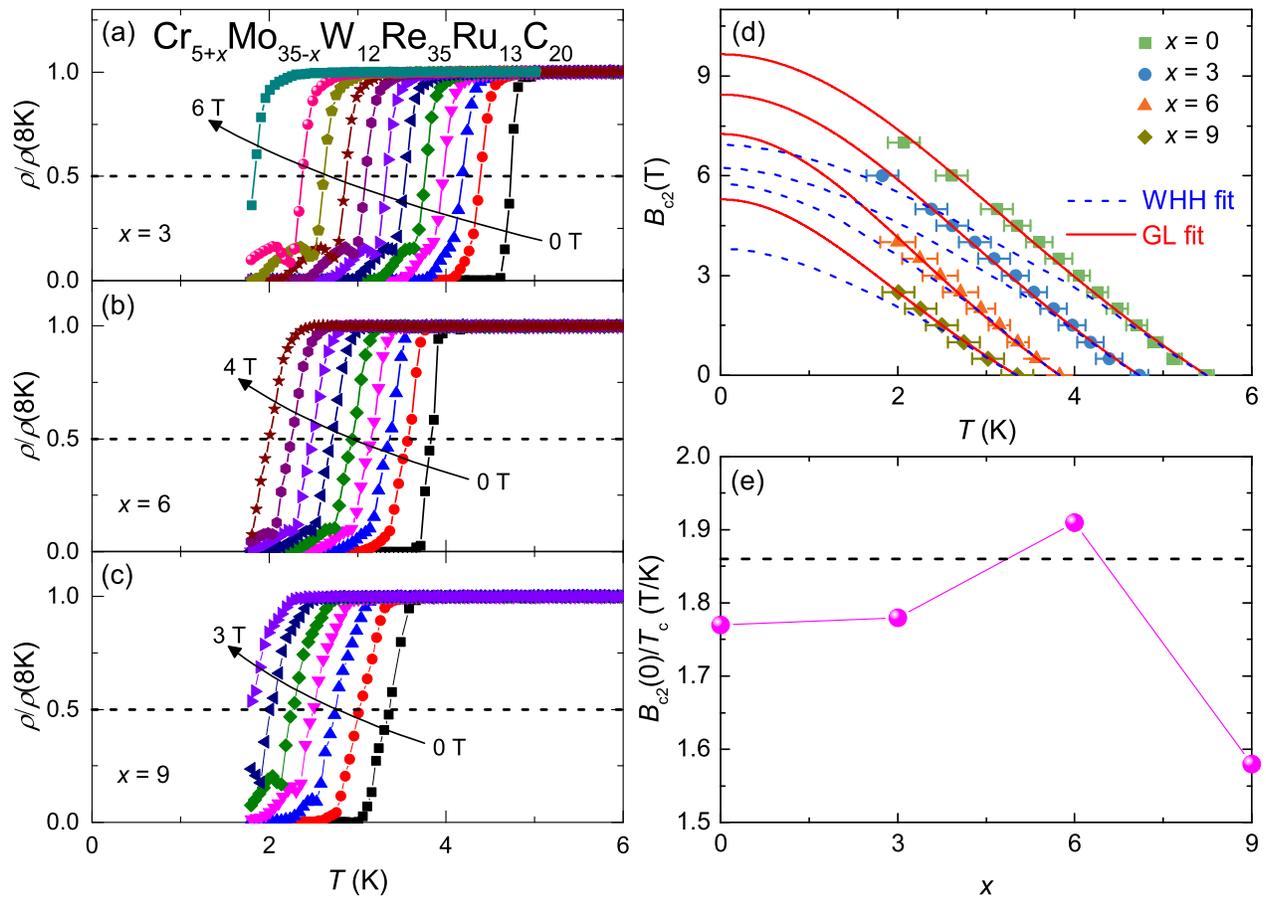}
	\caption{
		(a-c) Resistive transition under various magnetic fields for the Cr$_{5+x}$Mo$_{35-x}$W$_{12}$Re$_{35}$Ru$_{13}$C$_{20}$ HEAs with $x$ = 3, 6, and 9, respectively. In each panel, the arrow marks the field increasing direction. (d) Upper critical field versus temperature phase diagrams for the HEAs. The dashed and solid lines are fits by the WHH and GL models, respectively. (e) $x$ dependence of the $B_{\rm c2}$(0)/$T_{\rm c}$ ratio for the HEAs. The horizontal dashed line denotes the Pauli paramagnetic limit of $B_{\rm P}$(0)/$T_{\rm c}$ = 1.86 T/K.
	}
	\label{fig4}
\end{figure*}

\begin{figure*}
	\includegraphics*[width=16cm]{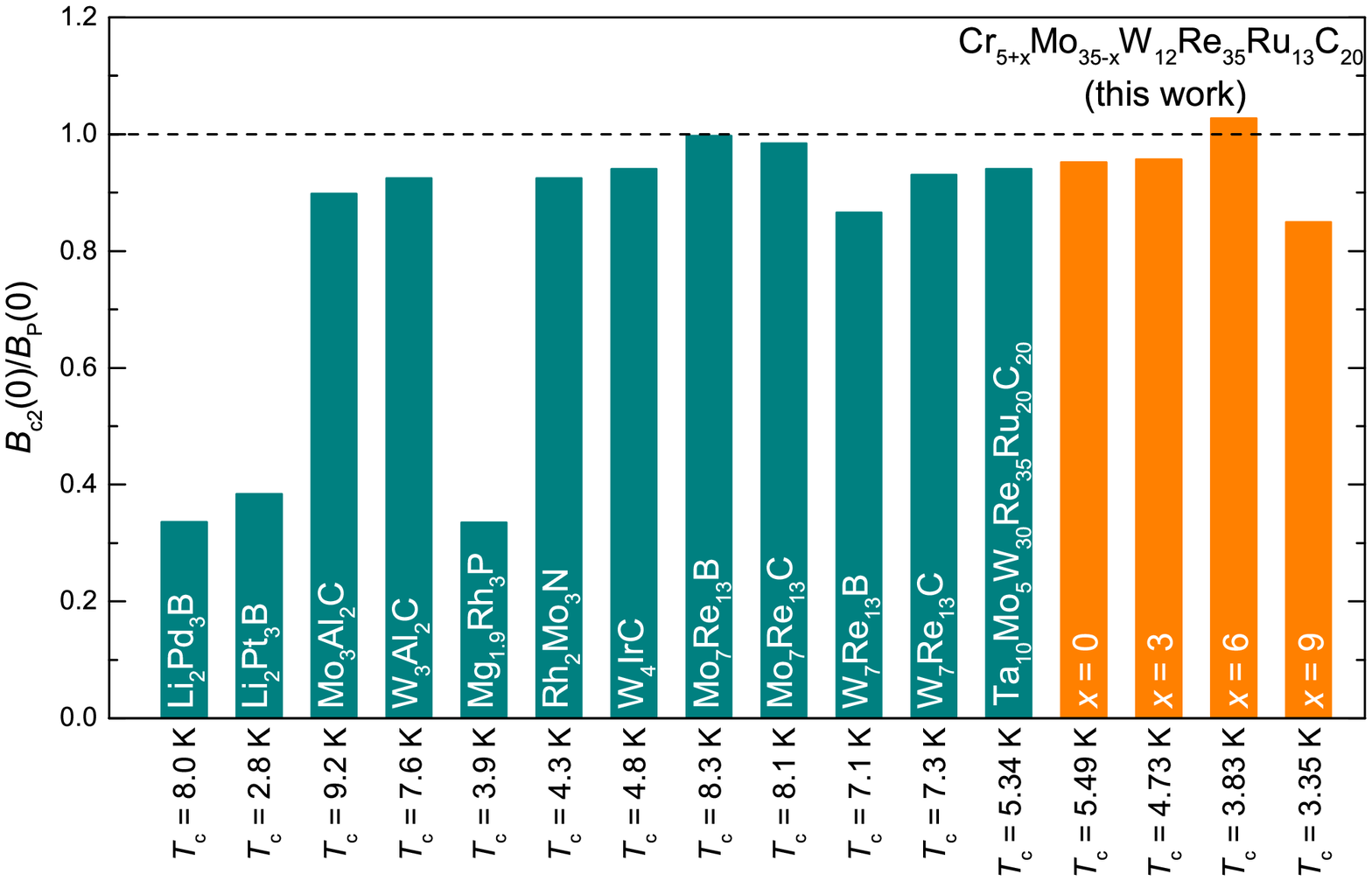}
	\caption{
	 $B_{\rm c2}$(0)/$B_{\rm P}$(0) ratios for various $\beta$-Mn type superconductors including the Cr$_{5+x}$Mo$_{35-x}$W$_{12}$Re$_{35}$Ru$_{13}$C$_{20}$ HEAs. The horizontal line represents the level of $B_{\rm c2}$(0)/$B_{\rm P}$(0) = 1.
	}
	\label{fig5}
\end{figure*}
The presence of local moments can induce spin-flip scattering, which breaks the Cooper pairs and leads to reduction in $T_{c}$.
This is indeed the case for the Cr$_{5+x}$Mo$_{35-x}$W$_{12}$Re$_{35}$Ru$_{13}$C$_{20}$ HEAs, as illustrated in the plot of normalized $T_{\rm c}$/$T_{\rm c0}$ against Cr fraction in Fig. 3(b).
One can see that the negative slope of $T_{\rm c}$/$T_{\rm c0}$ is close to that in La(Ru$_{1-x}$Cr$_{x}$)$_{3}$Si$_{2}$ \cite{li2016chemical}, and lies between V$_{1-x}$Cr$_{x}$ \cite{muller1959} and tetragonal Ta$_{10}$Mo$_{35-x}$Cr$_{x}$Re$_{35}$Ru$_{20}$ HEAs \cite{liu2021superconductivity}.
According to the Abrikosov-Gorkov (AG) theory \cite{abrikosov1960contribution} for magnetic pair breaking, the depression rate of $T_{\rm c}$/$T_{\rm c0}$ can be described by the universal relation
\begin{equation}
\ln(T_{\rm c}/T_{\rm c0}) = \Psi(\frac{1}{2})-\Psi(\frac{1}{2}+\alpha),
\end{equation}
where $\Psi$ is the digamma function and $\alpha$ is the pair breaking parameter.
The $T_{\rm c}$/$T_{\rm c0}$ data for the Cr$_{5+x}$Mo$_{35-x}$W$_{12}$Re$_{35}$Ru$_{13}$C$_{20}$ HEAs obeys well the AG formalism and the critical Cr fraction, at which $T_{\rm c}$/$T_{\rm c0}$ becomes zero, is estimated to be about 20.2\%. Note that this value is only half that for the Ta$_{10}$Mo$_{35-x}$Cr$_{x}$Re$_{35}$Ru$_{20}$ HEAs \cite{liu2021superconductivity}, implying a stronger pair breaking effect in the former compared with that in the latter.

We now turn the attention to the upper critical field $B_{\rm c2}$ of the Cr$_{5+x}$Mo$_{35-x}$W$_{12}$Re$_{35}$Ru$_{13}$C$_{20}$ HEAs.
Fig. 4(a)-(c) show the temperature dependencies of resistivity under various magnetic fields for $x$ = 3, 6, and 9, respectively.
For all cases, the resistive transition is suppressed toward lower temperatures as the field increases.
Nevertheless, a resistivity tail is observed in the intermediate to high-field region, which is reminiscent of that seen in the Ta$_{10}$Mo$_{5}$W$_{30}$Re$_{35}$Ru$_{20}$C$_{20}$ HEA \cite{xiao2022centrosymmetric}.
In fact, such anomaly is typical for $\beta$-Mn type superconductors containing carbon \cite{W7Re13X,zhu2022w}, which often precipitates at the grain boundaries and results in weak links between the superconducting grains.
As such, the $T_{\rm c}$ under field is determined by the same criterion as that at zero field and the resulting $B_{\rm c2}$ versus $T$ phase diagrams are summarized in Fig. 4(d).
The $B_{\rm c2}$($T$) data shows upward deviation from the Werthamer-Hohenberg-Helfand theory \cite{werthamer1966temperature}, especially at low temperature, but can be well fitted by the Ginzburg-Landau (GL) model
\begin{equation}
B_{\rm c2}(T) = B_{\rm c2}(0) \frac{1-t^{2}}{1+t^{2}},
\end{equation}
where $t$ = $T$/$T_{\rm c}$ is the reduced temperature.
Extrapolating the data to 0 K yields the zero-temperature upper critical field $B_{\rm c2}$(0) = 9.7 T, 8.4 T, 7.3 T, and 5.3 T for $x$ = 0, 3, 6, and 9, respectively. With $B_{\rm c2}$(0), the Ginzburg-Landau (GL) coherence length $\xi_{\rm GL}$(0) is found to vary from 5.8 nm to 7.9 nm based on the equation
\begin{equation}
\xi_{\rm GL}(0) = \sqrt{\frac{\Phi_{0}}{2\pi B_{\rm c2}(0)}},
\end{equation}
where $\Phi_{0}$ = 2.07 $\times$ 10$^{-15}$ Wb is the flux quantum.

In Fig. 4(e), we plot the ratio of $B_{\rm c2}$(0)/$T_{\rm c}$ as a function of $x$ for the Cr$_{5+x}$Mo$_{35-x}$W$_{12}$Re$_{35}$Ru$_{13}$C$_{20}$ HEAs.
Remarkably, $B_{\rm c2}$(0)/$T_{\rm c}$ attains large values above 1.58 T/K and exhibits a maximum of 1.91 T/K at $x$ = 6.
In fact, this maximum not only is slightly larger than the Pauli paramagnetic limit of $B_{\rm P}$(0)/$T_{\rm c}$ = 1.86 T/K \cite{PhysRevLett.9.266}, but also is the highest among $\beta$-Mn type superconductors.
This can be seen more clearly in Fig. 5, which presents a comparison of $B_{\rm c2}$(0)/$B_{\rm P}$(0) ratios between various superconductors of this family \cite{xiao2022centrosymmetric,togano2004superconductivity,badica2005superconductivity,karki2010structure,ying2019superconductivity,iyo2019superconductivity,wei2016r,W7Re13X,kawashima2006superconductivity,zhu2022w}.
Indeed, compared with nonmagnetic $\beta$-Mn type carbide superconductors, the Cr$_{5+x}$Mo$_{35-x}$W$_{12}$Re$_{35}$Ru$_{13}$C$_{20}$ HEA with $x$ = 6 has a considerably smaller $T_{\rm c}$ yet a obviously larger $B_{\rm c2}$(0)/$B_{\rm P}$(0) ratio of $\sim$1.03.
Also, the $B_{\rm c2}$(0)/$B_{\rm P}$(0) ratios of the HEAs with $x$ = 0 (0.95) and 3 (0.96) are slightly larger than those of Ta$_{10}$Mo$_{5}$W$_{30}$Re$_{35}$Ru$_{20}$C$_{20}$ (0.94) and W$_{4}$IrC (0.94) with similar $T_{\rm c}$ values.
At the highest $x$ of 9, however, the $B_{\rm c2}$(0)/$B_{\rm P}$(0) ratio of the Cr$_{5+x}$Mo$_{35-x}$W$_{12}$Re$_{35}$Ru$_{13}$C$_{20}$ HEA decreases to 0.85, which turns out to be the smallest among $\beta$-Mn type carbide superconductors.

For noncentrosymmetric superconductors, the ratio of $B_{\rm c2}$(0)/$B_{\rm P}$(0) might be an indication of the contribution from the spin-triplet pairing.
In this regard, it seems that the spin-triplet pairing component is enhanced within a certain range of Cr content.
However, this is unlikely since the spin-triplet pairing is unfavored in strongly disordered systems such as HEAs.
Instead, it is prudent to note that $B_{\rm c2}$(0)/$T_{\rm c}$ is proportional to $\gamma$$\rho_{\rm N}$ in the dirty limit \cite{PhysRevB.19.4545}, where $\rho_{\rm N}$ is the normal-state resistivity value just above $T_{\rm c}$.
Hence it is reasonable to speculate that the increases in both $\rho_{\rm N}$ and $\gamma$ are responsible for the enhancement of $B_{\rm c2}$(0)/$T_{\rm c}$ in Cr$_{5+x}$Mo$_{35-x}$W$_{12}$Re$_{35}$Ru$_{13}$C$_{20}$ HEAs with $x$ $\leq$ 6.
This is reminiscent of the cases in (V$_{0.5}$Nb$_{0.5}$)$_{3-x}$Mo$_{x}$Al$_{0.5}$Ga$_{0.5}$ HEAs \cite{wu2020polymorphism} and several superconducting amorphous transition metal alloys \cite{tenhover1981upper}.
The decrease of $B_{\rm c2}$(0)/$T_{\rm c}$ at a higher $x$ = 9 may be related to magnetic interaction of Cr, and the clarification of its orgin is definitively of future interest, in particular given that the interplay between magnetism and superconductivity in disordered noncentrosymmetric systems remains poorly understood.

In summary, we have synthesized and characterized a series of new Cr-bearing Cr$_{5+x}$Mo$_{35-x}$W$_{12}$Re$_{35}$Ru$_{13}$C$_{20}$ HEAs with 0 $\leq$ $x$ $\leq$ 9. In the whole range of $x$, the HEAs are found to crystallize in the noncentrosymmetric cubic $\beta$-Mn type structure and exhibit bulk superconductivity. With increasing $x$, the lattice constant shrinks and $T_{\rm c}$ is suppressed from 5.49 K to 3.35 K due to the magnetic pair breaking caused by Cr moments.
For all these HEAs, the $B_{\rm c2}$(0) is comparable to the $B_{\rm P}$(0) and the ratio of $B_{\rm c2}$(0)/$B_{\rm P}$(0) achieves a maximum of $\sim$1.03 for $x$ = 6, which is among the highest for $\beta$-Mn type superconductors. Our results call for further exploration of noncentrosymmetric HEA superconductors containing magnetic elements, which may help to better understand the interplay between magnetism, lack of inversion symmetry and superconductivity in strongly disordered alloy systems.\\

\section*{ACKNOWLEDGEMENT}
We acknowledge financial support by the foundation of Westlake University and the Service Center for Physical Sciences for technical assistance in SEM measurements. The work at Zhejiang University is supported by the National Natural Science Foundation of China (12050003).

\end{document}